\documentclass[a4paper,11pt,oneside]{article}
\usepackage{graphicx}				
\usepackage[colorlinks=true,linkcolor=green,anchorcolor=green,citecolor=green,filecolor=green,menucolor=green,urlcolor=blue,breaklinks=true,pdfhighlight=/P,pdfmenubar=true,pdftoolbar=true,pdfpagelabels=true,pdfstartpage=1,pdfstartview=FitV,pdftitle={Applications of the Dynamic Distance Potential Field Method},pdfsubject={Applications of the Dynamic Distance Potential Field Method},pdfauthor={Tobias Kretz},pdfcreator={Tobias Kretz},pdfproducer={Tobias Kretz},pdfkeywords={Pedestrian Traffic, Crowd Dynamics, Route Choice, Quickest Path, F.A.S.T. Model}]{hyperref}	
\usepackage[numbers,sort&compress]{natbib}
\usepackage{hypernat}	
\usepackage[american]{babel}
\begin{document}
\newlength{\figurewidth}\setlength{\figurewidth}{0.618\textwidth}
%
\title{Applications of the Dynamic Distance Potential Field Method}
\author{Tobias Kretz \\ \\
PTV Planung Transport Verkehr AG\\
Stumpfstra{\ss}e 1\\
D-76131 Karlsruhe\\ \\
\tt{Tobias.Kretz@ptv.de}}

\maketitle

\begin{abstract}
Recently the dynamic distance potential field (DDPF) was introduced as a computationally efficient method to make agents in a simulation of pedestrians move rather on the quickest path than the shortest. It can be considered to be an estimated-remaining-journey-time-based one-shot dynamic assignment method for pedestrian route choice on the operational level of dynamics. In this contribution the method is shortly introduced and the effect of the method on RiMEA's test case 11 is investigated.
\end{abstract}

\section{Introduction}
In traffic science a lot of effort is put into work discussing dynamic assignment methods: trying to find proofs on the existence of an equilibrium and/or implementing it in an efficient manner for the simulation of real-world networks. This effort is justified by the conviction that Wardrop's first principle ``The journey times in all routes actually used are equal and less than those which would be experienced by a single vehicle on any unused route." \cite{Wardrop1952} is at least a good approximation for real world traffic flow distribution. 

Compared to this strong emphasize of (estimated remaining) journey time in vehicular traffic science as a determinant of route choice, explicit discussion of the influence on the dynamics of pedestrians \cite{Schadschneider2009,Schadschneider2009b} is surprisingly scarce. Although evidently in case of emergency it's their egress {\em time} that occupants desire to minimize and not their egress path length, models of pedestrian evacuation dynamics are often constructed such that various influences change the basic direction of desired velocity, which is calculated as a gradient of the field of {\em distances} toward the exit and not a field of estimated remaining journey time. While this is not to say that pedestrian traffic always moves on quickest paths -- there are probably more occasions than in vehicular traffic when it deviates from it -- there are at least numerous situations, where it is close to travel time referred user-equilibrium or at least travel time is relevant for route choice.

Recently introduced dynamic distance potential fields \cite{Kretz2009,Kretz2009b} do not hold information on remaining journey time either, but -- on any spatial scale -- on the load of the remaining path and implicitly the possibility to detour jams. With this information it is possible to shift the agents' walking behavior toward user-equilibrium. 

In the two subsequent sections of this paper at first the method of dynamic distance potential fields is shortly sketched and then applied in a simulation of RiMEA's test case 11 \cite{Rimea2009}.

\section{Dynamic Distance Potential Field}
\begin{figure}[htbp]
  \center
	\includegraphics[width=0.618\textwidth]{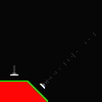}
	\caption{A visualization of the dynamic distance potential field. The green line is the exit, agents are located on the bright white spots, which by that have a large value of the dynamic distance potential field. The darker a spot is, the smaller is its dynamic potential field value. The gradient of the dynamic distance potential field enters into the calculation of the direction of desired motion.}
	\label{fig:shadow}
\end{figure}

Let's begin with a summary of the method: a dynamic distance potential field can be imagined as a ``shadow" (as shown in figure \ref{fig:shadow}) of agents cast ``upwards" the static potential, i.e. away from the destination. One can think of a destination area as a source of light sending light rays that travel under consideration of obstacles on approximate geodisces away from the destination area. When an agent or a group of agents throws a shadow, successive agents will try to avoid the shadow, as it is a hint for upcoming delays. A shadow is the longer the larger and denser a group of agents (i.e. a jam) is.

This functionality is achieved using flood fills, where a value of 1 is added, if a cell is unoccupied and some larger value $s_{add}$, if it is occupied by an agent. But the two simple flood fill methods (over common edges or common edges and corners) result in Manhattan or Chebyshev but not the desired Euclidean metric. In this sense these two methods produce large errors and unwanted artifacts in the movement. However, the errors can be reduced for once by combining the two methods and second by using as heuristic not the time dependent potential, but its difference to the potential calculated on the empty (unoccupied) geometry.

\begin{figure}[htbp]
  \center
	\includegraphics[width=0.22\textwidth]{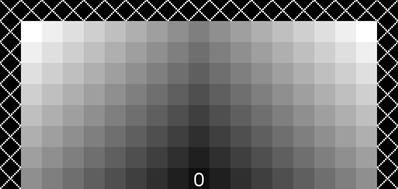}\hspace{5pt}
	\includegraphics[width=0.22\textwidth]{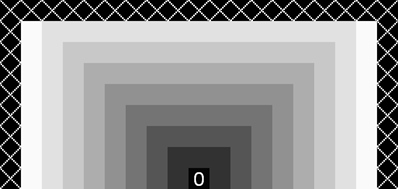}\hspace{5pt}
	\includegraphics[width=0.22\textwidth]{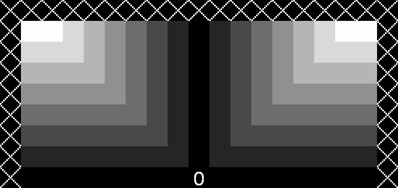}\hspace{5pt}
	\includegraphics[width=0.22\textwidth]{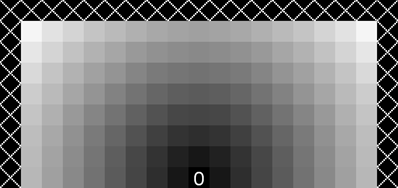}
	\caption{F.l.t.r.: Manhattan metric, Chebyshev metric, Minimum norm, and Variant 1 (which here is identical to Euclidean metric).}
	\label{fig:norms}
\end{figure}

For two mutually visible coordinates $(x_0,y_0)$ and $(x_1,y_1)$, Manhattan distance (i.e. a distance according to Manhattan metric) is
\begin{equation}
d^M=|x_1-x_0|+|y_1-y_0|=|\Delta x| + |\Delta y|
\end{equation}
and Chebyshev distance is
\begin{equation}
d^C=\max(|\Delta x|,|\Delta y|)
\end{equation}
From this the ``minimum distance" follows:
\begin{equation}
d^m=d^M-d^C=\min(|\Delta x|,|\Delta y|)
\end{equation}
and therefore the Euclidean distance is
\begin{equation}
d^E=\sqrt{(d^C)^2+(d^m)^2}=\sqrt{|\Delta x|^2+|\Delta y|^2}
\end{equation}
Combining the whole potentials cell by cell in this manner -- even if cells are not mutually visible -- will now be called ``method V1". All mentioned metrices and norms are depicted in figure \ref{fig:norms}. A discussion of the resulting errors of this method compared to Euclidean distance is given in \cite{Kretz2008c}.

Let's call the potential calculated on the empty (no agents) geometry $S_{V1}^0$ and the potential after a certain time step $S_{V1}(t)$. Then the heuristic influencing the motion of pedestrians is $S_{dyn}(t)=S_{V1}(t)-S_{V1}^0$.

This has been used and coupled to the F.A.S.T. model \cite{Kretz2006f,Kretz2006d,Kretz2007a,Kretz2008f,Kretz2009e} as an additional partial probability 
\begin{equation}
p_{dyn}=e^{-k_{Sdyn}S_{dyn}(t)},
\end{equation}
multiplied to the original probability that a cell is selected as destination cell and normalized accordingly
\begin{equation}
\hat{p} = \hat{N}\cdot p \cdot p_{dyn}.
\end{equation}

\section{RiMEA Test Case 11}
\begin{figure}[htbp]
  \center
	\includegraphics[width=\figurewidth]{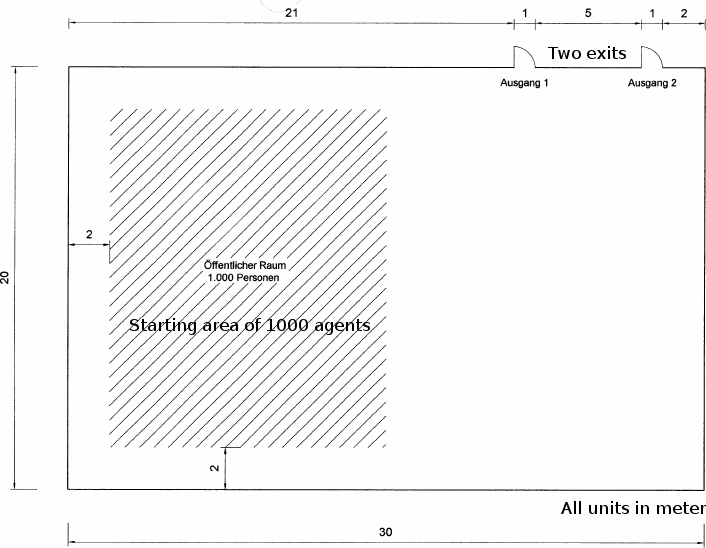}
	\caption{Test case 11 of RiMEA.}
	\label{fig:rimea11}
\end{figure}

In RiMEA's test case 11 a group of 1000 agents is meant to leave a room using two doors, which have different distance to the agents' starting area. The geometry of test case 11 is shown in figure \ref{fig:rimea11}. RiMEA prescribes an exit width of 1 m for the simulations. As space is discretized into cells of width 40 cm in the F.A.S.T. model, the exit width has been set to 80 cm. The parameters of the F.A.S.T. model were set to $k_S=1.0$ and any other $k=0$. The speeds were normal distributed (3.5 cells/s, $\pm$ 1.0 cell/s) with cut-offs at 1 and 4 cells/s. All results given are averages of 100 simulation runs.

\begin{figure}[htbp]
  \center
	\includegraphics[width=0.3\textwidth]{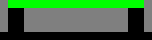} \hspace{10pt}
	\includegraphics[width=0.3\textwidth]{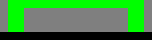} \hspace{10pt}
	\includegraphics[width=0.3\textwidth]{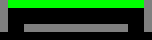}
	\caption{Three variants of geometric exit modelling. Agents are taken out of the simulation as soon as they reach the green area, obstacles are grey, free space black. Note that in any case there is only one destination area with one static floor field, such that the exit choice method proposed in \cite{Kretz2007a} is not made use of in the simulations for this contribution.}
	\label{fig:variants}
\end{figure}

In the description of the test case the shape and position of the actual destination area, i.e. the area from which agents are taken out of the simulation upon arrival, is not defined. Therefore three variants were tested, which are shown in figure \ref{fig:variants}.

Tables \ref{tab:times} and \ref{tab:right} show in numbers the effect the dynamic distance potential field has. Simulated evacuation times shrink as the agents distribute more equally on the two exits. As a by-observation the effect of the exit geometry becomes obvious.

\begin{table}
\center
\begin{tabular}[htbp]{c||c|c||c|c}
Exit geometry  & $T$ w/o DDPF &  $T$ with DDPF & $T_i$ w/o DDPF & $T_i$ with DDPF  \\ \hline
    1          &    588.8     &       474.9    &  270.4         &  229.1         \\
    2          &    447.9     &       289.1    &  212.6         &  143.5     \\
    3          &    539.5     &       400.4    &  248.0         &  191.3    \\
\end{tabular}
\caption{Total times $T$ and average individual egress times $T_i$ without and with dynamic distance potential field. In the latter case $k_{Sdyn}=1.0$ and $s_{add}=10$.}
\label{tab:times}
\end{table}

\begin{table}
\center
\begin{tabular}[htbp]{c||c|c}
Exit geometry  & w/o DDPF   &  with DDPF   \\ \hline
    1          &  407.0     &   495.7      \\
    2          &  198.2     &   483.4      \\
    3          &  338.9     &   492.3      \\
\end{tabular}
\caption{Number of agents that leave through the right exit. }
\label{tab:right}
\end{table}

\begin{figure}[htbp]
  \center
	\includegraphics[width=\figurewidth]{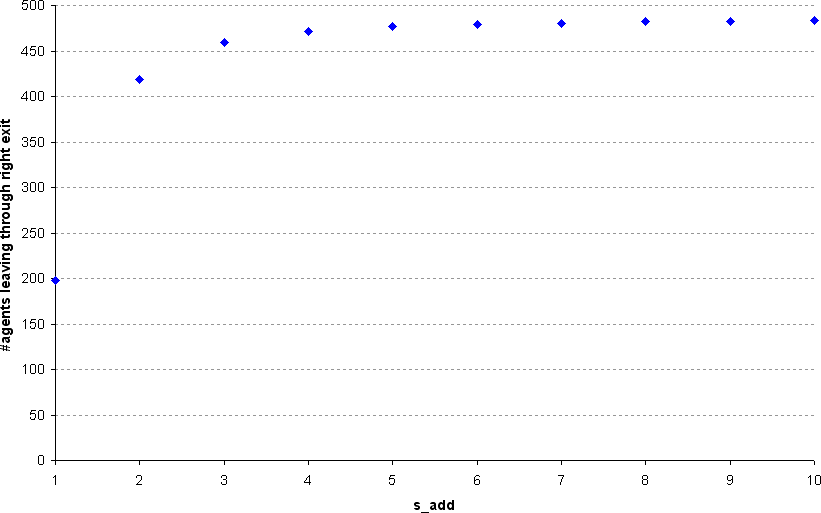}
	\caption{Number of agents leaving through the right exit in dependence on parameter $s_{add}$ (exit geometry 2).}
	\label{fig:saddright}
\end{figure}

Ammending table \ref{tab:right} figure \ref{fig:saddright} shows for exit geometry 2 the dependence of the number of agents leaving through the right exit in dependence on parameter $s_{add}$.

\begin{figure}[htbp]
  \center
	\includegraphics[width=0.45\textwidth]{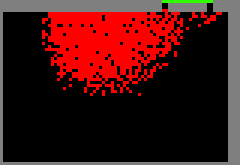} \hspace{10pt}
	\includegraphics[width=0.45\textwidth]{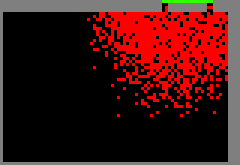}
	\caption{Situation after 100 seconds without (left) and with (right) dynamic distance potential field.}
	\label{fig:spatial}
\end{figure}

The difference becomes even more obvious, if one looks at the situation after 100 seconds in figure \ref{fig:spatial}. One gets the impression that without dynamic distance potential field the agents are either only in small numbers ``aware" of the existence of the right exit or are reluctant to walk five extra meters. This is totally different with dynamic distance potential field. Here it's almost difficult to distinguish the shape of the crowd from a geometry, where there is only one exit placed between the two actually existing ones.

\section{Summary and Conclusions}
The dynamic distance potential field method has shortly been described. It is a computationally efficient heuristic method to mimic the effects of estimated remaining journey time on route choice on the operational level of pedestrian dynamics. For fundamental reasons, but also to keep computation times small, deviations from real remaining journey time and limited errors in the metric of the field are accepted.

The method has been applicated in a simulation of RiMEA test case 11. This led to the agents utilizing the exit 5 m further away almost as well as the nearby exit, which could best be seen in a snapshot from the simulation's animation.

As long as there is no experiment for test case 11 or an observation that resembles test case 11, it's not safe to say that the dynamic distance potential field makes the simulation in this case more realistic. But with the two parameters that control the method a planner has the flexibility to model almost any point between strong preference for shorter walking paths and strong preference for shorter walking times, respectively shorter evacuation times.

%
\nocite{_ACRI2006,_ACRI2008,_PED2005,_PED2008,_Bazzan2009,_Enzy2009}
\bibliographystyle{utphys_quotecomma}
\bibliography{025_TGF09}
%
\end{document}